\documentclass{aastex}
\usepackage{emulateapj5}

\slugcomment{Submitted to ApJL}

\shorttitle{Near-infrared coronagraphy of TW Hya}
\shortauthors{Trilling et al.}

\begin{document}

\title{Near-infrared
coronagraphic
imaging of the circumstellar disk around TW Hydrae}

\author{D. E. Trilling\altaffilmark{1,2},
D. W. Koerner\altaffilmark{1},
J. W. Barnes\altaffilmark{2,3},
C. Ftaclas\altaffilmark{4},
R. H. Brown\altaffilmark{2,3}
}

%\affil{University of Pennsylvania, David Rittenhouse Laboratory,
%209 S. 33rd Street, Philadelphia, PA 19104}

%\email{trilling@hep.upenn.edu}

%\author{Ray Jayawardhana}
%\affil{Department of Astronomy, University of California,
%Berkeley, CA 94720}

%\author{Jason W. Barnes\altaffilmark{1}}
%\affil{Lunar and Planetary Laboratory, University
%of Arizona, Tucson, AZ 85721}

%\author{Christ Ftaclas}
%\affil{Department of Physics, Michigan Technological
%University, 1400 Townsend Dr., Houghton, MI 
%49931}

%\and

%\author{Robert H. Brown\altaffilmark{1}}
%\affil{Lunar and Planetary Laboratory and Steward
%Observatory, University
%of Arizona, Tucson, AZ 85721}

\altaffiltext{1}{University of Pennsylvania, David Rittenhouse
Laboratory, 209 S. 33rd Street, Philadelphia, PA 19104;
trilling@hep.upenn.edu}
%\altaffiltext{2}{trilling@hep.upenn.edu}
\altaffiltext{2}{Visiting Astronomer at the
        Infrared Telescope Facility which is operated
        by the University of Hawaii under contract
        to the National Aeronautics and Space
        Administration}
\altaffiltext{3}{Lunar and Planetary Laboratory, University
of Arizona, Tucson, AZ 85721}
\altaffiltext{4}{Department of Physics, Michigan Technological
University, 1400 Townsend Drive, Houghton, MI 49931}
%\altaffiltext{5}{Steward Observatory, University of Arizona,
%Tucson, AZ 85721}

\begin{abstract}
We present ground-based near-infrared (H-band) imaging of the circumstellar
disk around the nearby classical T~Tauri star TW~Hydrae.
The scattered light image shows a face-on disk
with radius 
4~arcseconds
(corresponding to 225~AU)
and a morphology which agrees
with recent
images from the Hubble Space
Telescope and the Very Large Array.
The best fit power-law for the disk's
radial surface brightness profile
obeys the law $r^{-3.3\pm0.3}$.
%some unresolved
%disk structure is also apparent in
%addition to the power law.
We use
our image and published continuum flux densities
to derive
properties of the disk with a simple model of
emission from an optically thick, flat disk.
%Our model freely parameterizes disk mass
%and inner radius and assumes radial power-law
%functions for temperature and surface density
%($T \propto r^{-q}$ and $\Sigma \propto r^{-p}$)
%and a grain
%opacity which varies as a power law in frequency ($\kappa \propto
%\nu^{-\beta}$).
The best-fit values for disk
mass and inner radius are 0.03~$M_{\odot}$
and 0.3~AU; best-fit values for
temperature, density, and grain opacity
power law exponents 
($q$, $p$, and $\beta$)
are~0.7,~1.3, and~0.9, respectively. These properties are
similar to those of disks around
%detected in mm-wave
%surveys of
classical T~Tauri stars located in more distant
molecular clouds. %Additionally, the mass is
%similar to the minimum required to make up our own solar system.
Because of TW~Hydrae's nearby location and pole-on orientation,
it is a uniquely favorable object for future studies of radial
disk structure at the classical T~Tauri stage.
\end{abstract}


\section{Introduction}
The formation of the solar system from a nebula of gas and dust is
strongly supported by observations of disks around young stars
\citep{tetons}.
Observations of these protoplanetary disks
are largely confined
to
star-forming
clouds at distances greater than 140~pc.
In these regions, the 
majority of classical T~Tauri stars (cTTs)
are found to have disks
with masses in the range
10$^{-2}$~to~10$^{-4}$~$M_{\odot}$
\citep{beck90,am94}.
Aperture
synthesis mapping of circumstellar gas has established
that these disks have radii of a few 100 AU
\citep{dwkais95,dutrey96}, %,mannings},
in agreement with dimensions derived from
HST images of young disks \citep{odell,padgett}. %,stapelfeldt}.

The discovery of a nearby lone classical T~Tauri star by \citet{rucinski} 
affords a unique opportunity
for high spatial resolution studies of a 
gas-rich disk.
TW~Hydrae (TW~Hya, 
K7Ve, $d\approx56$~pc) is the eponymous member of a nearby 
group of young stars 
\citep{kastner97,webb}.
The members of the TW~Hydrae
Association have ages of only a few to ten million years
despite their distance from a molecular cloud
\citep{jensen,webb}.
TW~Hya (the star)
stands in contrast to its namesakes
by
virtue
of its
large H$\alpha$~equivalent width (220~\AA) \citep{kastner97},
CO~emission \citep{weintraub,zuck,kastner97},
%(observed by Mundy, reported in 
%\citet{weintraub}; see also \citet{zuck},
%\citet{kastner97}),
and 
molecular hydrogen \citep{wein2000}.
Furthermore, 
the considerable 1.1~and 7~mm flux densities of 
874~mJy and 8~mJy, respectively \citep{weintraub,wilner} --
comparable, after distance
scaling, to flux densities of cTTs at much larger
distances --
argue for an optically thick, massive ($\sim10^{-2}~M_{\odot}$)
circumstellar disk.
This disk appears to be still actively 
accreting onto the star, as inferred from H$\alpha$
emission \citep{muzerolle}.
In short,
TW~Hya shows all the signs of being a classical T~Tauri
star with a surprisingly advanced age.

Recently, \citet{krist} (hereafter K00) and \citet{alycia} reported the
detection of a face-on (radius~$\approx$~3.5\arcsec ) disk around TW~Hya in
HST~images obtained with WFPC2 and NICMOS,
respectively. %, after subtraction of stellar point spread
%functions.
\citet{wilner}
resolve 
(radius~$\gtrsim$~2\arcsec) thermal emission from circumstellar dust
around TW~Hya with 7~mm VLA~observations.
In this Letter, we present the first ground-based,
near-infrared image of the circumstellar disk
around TW~Hya, and combine this data
with thermal emission data  
to derive disk properties with a simple disk model. We find disk
properties that are similar to those of disks
around other cTTs.
The nearby TW~Hya %therefore
provides the
best known opportunity to study the earliest
stages of potentially planet-forming systems.

\section{Observations and Results}

We observed TW~Hya on 7~Apr~2000~(UT) from NASA's
Infrared Telescope Facility (IRTF) on Mauna 
Kea. We used CoCo, the Cold Coronagraph,
to block the light from the central star and reduce
sidelobe intensity to
allow imaging of the circumstellar disk;
CoCo is used 
as a front end to the 
IRTF facility camera
NSFCAM \citep{rayner,shure}.
CoCo is a cryogenically cooled coronagraph
with gaussian apodized focal plane masks and 
a cooled articulatable pupil plane mask \citep{wang,toomey}.
We used a mask
whose half-power (50\%~transmission)
diameter is~1$\farcs$86.
A series of eight exposures of TW~Hya, typically 300~seconds
long, were interspersed with similar ones of nearby
sky. Total on-source integration time was 2220~seconds.
We also obtained six 20-second exposures of the nearby
G~star SAO~202496 to measure the point spread
function (PSF); these observations were made immediately
after the TW~Hya observations.
SAO~202496, which has a similar
H~magnitude to TW~Hya ($\Delta$mag~$\lesssim$~0.5)
and a sky separation of 6.4~degrees,
is 
well suited as a PSF~comparison star for use
in analysis of the TW~Hya images.
We observed both stars in H~band
($\lambda_c$~=1.62~\micron ,
$\Delta\lambda$~=~0.28~\micron );
seeing was consistently 0$\farcs$8 at 
airmasses 1.7-1.8~for TW~Hya and 
1.8~for SAO~202496.
We also observed the photometric flux standard SAO~205948
\citep{elias}
at similar airmasses.
%for photometric calibrations.

All exposures were sky-subtracted and flat-fielded.
Each frame of SAO~202496 was aligned, scaled,
and subtracted from each TW~Hya frame.
%for 
%a total of 48~independent measurements. 
Alignment and scaling of the target and PSF~frames
were accomplished by 
minimizing
the least squares difference ($\chi^2$)
between target and PSF~frames while varying
amplitude scaling and position offsets. 
The least squares difference was calculated in
a 
region
4~to 6.5~arcsec from the center of the frame.
Image shifting in
%with spatial resolution
steps of 0.1~pixel
was carried out in the Fourier
domain. The amplitude was varied
in step sizes corresponding to approximately
1\%~of the best fit value.
We carried out
this same procedure
%aligning, scaling, and subtracting
on sets of two
independent
exposures
of the PSF~star to ensure that PSF$-$PSF~results produced a null
result 
and
to characterize artifacts and residuals.
Circular emission 3-4~arcsec
in radius 
was detected in all
%48~independent 
48~combinations of target$-$PSF frames with a signal to noise ratio of 
several hundred compared to the background~RMS
and 80~with respect to residuals in
the PSF$-$PSF~subtraction.
Because CoCo rejects 90\%-95\%
of the light from the central star,
scattered light from the circumstellar disk
itself dominates 
from  
2\arcsec\ radius out to the edge of the disk (see below).

Figure~1a shows our best 
image
of the nebulosity around TW~Hya.
A circular halo of size r=4\arcsec\
(5$\sigma$~detection limit)
is clearly apparent. %, in agreement
%with the image of K00.
The peak intensity of the emission is 6.8$\pm$1.1~mJy/arcsec$^2$
at a radial distance of 0$\farcs$94~from the star,
the 
smallest radial distance we can see using
this focal plane mask.
The reported uncertainty comprises a conservative
estimate of errors.
The largest source of error is imperfect centering
of the stars behind the coronagraphic mask.
This effect results in azimuthal variations of 
approximately 15\%~in the final subtracted
image when averaged over complete
quadrants; this is estimated by assuming
that the circumstellar emission is axisymmetric
and centered on the star.
By contrast, the
1$\sigma$~pixel-to-pixel RMS~noise in the image background 
is 10~$\mu$Jy/arcsec$^2$, corresponding to 
an error of much less than~1\%. (This background
noise is dominated by noise from the scaled
PSF~star background.)
The conservative nature of our photometric 
error estimate is illustrated by subtractions
of independent observations of the PSF~star
(Figure~1b).
Here we show the difference between two
independent measurements of SAO~202496;
the residuals have a maximum above-background
emission at  
just over 1\arcsec\ with an average peak
surface brightness
of 0.1~mJy/arcsec$^2$.
These images are subject to the same centering
errors as the PSF-subtracted image of TW~Hya,
and the residuals seen in the PSF$-$PSF image
can represent the error due to imperfect centering.
The total integrated flux density from the circumstellar
emission around TW~Hya is 21.6$\pm$3.5~mJy;
by comparison, the total integrated flux density
of the residuals in the
PSF$-$PSF image is around 0.25~mJy.
The fidelity of the final TW~Hya image is bolstered
by the comparison displayed in Figure~1c.
Here we show the K00~WFPC2~I~band image in grayscale
overlain 
with contours from our TW~Hya image (Figure~1a).
In radial extent and orientation, there
is a very clear and convincing match between
the two.

In 
Figure~\ref{rad}
we show azimuthally averaged radial profiles
of TW~Hya and PSF~star images before PSF~subtraction
(thin solid lines) as well
as the radial profile of the final,
PSF-subtracted image of the 
circumstellar emission around TW~Hya
(thick solid line).
We also show the azimuthally averaged
radial profile of the PSF$-$PSF residual
(thick dotted line).
This figure illustrates a point made
above: outside of~2\arcsec ,
a circumstellar excess is seen in the
TW~Hya image compared to the PSF~star image
before PSF~subtraction from either image.
This excess extends to around 4\arcsec\
in the fully-reduced (PSF-subtracted)
data (thick solid line in Figure~\ref{rad}),
corresponding to an outer radius
of 225~AU at the distance 
of TW~Hya. 
A single power-law fit to the radial profile follows
the law~$r^{-3.3\pm0.3}$ (thick dashed line
in Figure~\ref{rad}),
although the data does not precisely follow
a single power law (see Figure~\ref{rad}).
Although we find scant evidence for the 
multiple power laws derived by K00,
we do not have the spatial resolution 
which K00 had, and 
our
results are in good agreement with the
average slope 
($\sim$-3.5) of the K00 broken
power law. % (see below).
Our data may show 
that the radial profile of the disk
becomes steeper outside of 
200~AU, in rough
agreement with K00;
this level is, however, close to the background level.

\section{Modeling and discussion \label{disc}}

The 1.1~mm continuum flux density from TW~Hya
(874$\pm$54~mJy)
implies a total circumstellar mass (gas+dust) of order
10$^{-2}~M_{\odot}$ \citep{weintraub}.
For micron-sized grains,
this implies
a total dust grain cross sectional area
around $10^7$~AU$^2$,
more than large enough
%1~and 10~$\mu$m grain sizes, respectively,
%this implies a total dust grain cross section
%of
%around~10$^7$ and~10$^6$~AU$^2$,
%more than enough area
to intercept all stellar light if 
oriented in a spherical shell with radius
similar to the disk size.
Additionally, the fractional disk luminosity is
$\sim$0.3 \citep{zuck} but the 
ratio
of disk reflected light to stellar light is only~$\sim$$10^{-2}$
at H~band.
Both statements imply that
the optical depth in the disk midplane must be quite high. This precludes
the use of a model which derives temperature structure under the assumption
of direct thermal equilibrium between stellar radiation and individual
grains (e.g., \citet{backman}).
Instead, the temperature structure may reflect
a variety of
effects, including viscous internal
heating, re-processing of stellar
radiation, and super-heating in a surface layer
(e.g., \citet{lb,chiang}).
Without {\it a priori} knowledge of the relative
magnitudes of these influences, we find it simplest
to freely parameterize the temperature structure
in a fit to the spectral energy distribution (SED)
as in \citet{beck90}.

We derive properties of the TW~Hya circumstellar
disk by using a simple, two-dimensional model which 
incorporates both our imaging data and the thermal
continuum flux densities %(SED)
shown in Figure~\ref{sed}.
Our best disk model is also shown in Figure~\ref{sed}.
Following \citet{beck90}, we
assume power-law radial profiles for the temperature and surface
density, $T = T_1(r)^{-q}$
and 
$\Sigma = \Sigma_0(r/r_0)^{-p}$, where
$r$~is in~AU, $T_1$~is the temperature at 1~AU,
and 
$\Sigma_0$ (the surface density at $r_0$)
is couched in terms of the disk mass, $M_{disk}$, by integrating
over the disk surface area. The grain emissivity is $\kappa =
0.1(\nu/10^{12}$~Hz)$^\beta$ cm$^2$ g$^{-1}$.
%The temperature scaling
$T_1$~and
%and power-law index
$q$~are found to be 125~K and~0.7,
respectively, by fitting points
at
infrared wavelengths where the 
disk radiation is optically thick; $T_1\approx$~125~K is
similar to results for other cTTs \citep{beck90}.
An inner radius~($R_0$) of 0.3~AU is required to fit the 12~$\mu$m
point. The face-on orientation and outer radius of 225~AU are
indicated by the image in Figure~1a. We vary the remaining parameters
($M_{disk}$,~$p$, and~$\beta$) over a grid of appropriate values
while minimizing the reduced~$\chi^2$ difference between
model and data. Our derived best-fit
values are $M_{disk}$~=~0.03 $M_{\odot}$,
$p=1.3$,
and $\beta$~=~0.9.

Typical $q$~values for~cTTs 
($\sim$0.5; \citet{beck90})
can indicate back-warming from an envelope
or reprocessing by a geometrically flared
disk
\citep{natta,chiang}.
Our larger value for $q$ (0.7) is closer to the value expected for
either a standard viscous accretion or 
geometrically flat reprocessing disk, perhaps
suggesting the
disk flattening implied for
some
transitional
sources \citep{miyake}.
Radial brightness power law exponents
for geometrically flat to fully flared disks
are expected to range from~-3 to~-1.8.
Our power-law
fit exponent is -3.3$\pm$0.3, consistent 
with the average K00 exponent of $\sim$-3.5. 
Steeper indices than~-3 might result from
regions of the disk which are shadowed from
the star. 
The higher angular
resolution K00 observations
yield regions with power-law exponents between
-0.5~to~-5.9,
perhaps suggesting
%show that the
%I~band surface brightness exponent
%varies with stellocentric distance in I~band,
%takes values of -1.7,~-5.0,~-0.5, and~-5.9
%in annuli of increasing stellocentric 
%distance.
%Thus, the K00 data 
%may suggest
the kind of ripples
which can exist for passively
heated disks \citep{ec_phd,dullemond}.

The inner disk hole ($R_0=0.3$~AU) 
implied by our model is 
required
by the deficit 
of disk-derived
flux at 12~\micron\
compared to that at 18~\micron .
However,
more sophisticated models of %flared,
face-on disks can produce a relative decrease
at 12~\micron\ 
without an inner disk hole \citep{chiang}; our model
simply requires an inner disk hole
as a substitute for more complex
radiative transfer calculations.
Future work includes applying these more sophisticated
models to our 
TW~Hya data.

%The
%temperature and density profiles
%we derive from modeling the disk continuum
%emission agree very well with those obtained
%by K00 from modeling the scattered-light
%intensity profile.
%Our continuum emission model
%results in 
We derive
a temperature power law index
of $q=0.7$ from the disk continuum emission,
consistent 
with the 
average K00 scattered-light
disk scale height 
radial power law
exponent of $\sim$1.15
\citep{kh87}.
Similarly, both we and K00 derive
(average) density power law indices of $p=1.3$.
Our value for~$p$ is not narrowly constrained,
and values between
1.1~and~1.5 all provide acceptable fits
to the continuum data; however,
our~$p$
is better constrained than is typical for
SED fits since the principal degeneracy is with
outer disk radius, which is given here 
by our image.
K00 likewise found their
density exponent not well constrained. 
A value of $p=0.75$ is the theoretical expectation
for steady-state accretion \citep{lb},
although \citet{bell} incorporate effects 
such as disk opacity and shadowing
and find
complicated radial density distributions
with effective average~$p$ closer to unity.

Our derived $M_{disk}$~of~0.03~$M_{\odot}$
and $\beta$~of~0.9 are 
narrowly constrained 
by the long wavelength (0.8,~1.1,~and 7~mm) data;
\citet{wilner}
found~$M_{disk}=0.03~M_{\odot}$
and~$\beta=1.0$.
These derived $\beta$~values % of $\beta$
differ from $\beta$
for the ISM ($\sim$2), but agree with
that found for dust around other cTTs
\citep{bs91,dutrey96}.
%mathis
This effect has been explained as
the result of aggregation of dust grains into larger
disk particles \citep{beckppiv}.

The age derived for TW~Hya,
3~to~$10\times10^6$~yr
\citep{jensen,webb},
is somewhat older than 
the ages typically reported for 
other cTTs with similar disks,
10$^5$~to
$5\times10^6$~yr \citep{beck90}.
%typical%ly reported
%for other
%cTTs with similar disks;
%most cTTs with strong mm-wave
%continuum excess have ages between 10$^5$~and~5$\times$10$^6$~yr
%\citep{beck90}.
It is therefore surprising
that TW~Hya possesses
youthful disk properties,
even though it exhibits a stellar age 
at the upper end of this range and
disks around other Association members 
are considerably more evolved than the TW~Hya
disk \citep{dwk98,rayjay98,rayjay99a,rayjay99b,schneider,dwk2000}.
However,
because there can exist a wide variation in initial disk masses and
lifetimes 
(e.g., \citet{ob95}) and
because 
ages for~cTTs in star-forming regions
have great uncertainties (due to poorly ascertained distances
and %uncertainties in
extinction corrections),
we suggest that disk properties should be taken as
the key indicator of a disk evolutionary state which is
only approximately correlated with
HR~diagram-derived 
stellar ages.
In this view, TW~Hya is
especially useful as the closest example of a protoplanetary
disk in the 
classical T~Tauri phase
and with a mass similar to the minimum
mass solar nebula.
Since disk properties at this
stage are likely to hold great importance
for theories about the origin of planetary systems,
TW~Hya's nearby location and face-on orientation
make it ideal for 
future high-resolution
studies of %radial structures in
T~Tauri
disks.

\acknowledgments
We thank
%Lars Bergknut, George Koenig, Paul Jensen,
%and the 
the IRTF day crew and telescope operators for their
invaluable assistance.
We thank R. Jayawardhana for help with the 
planning, execution, and interpretation of these observations.
We thank Z. Wahhaj, H. Klahr, and E.
Chiang, and J. Krist for useful conversations,
%J. Krist for discussion of the HST/WFPC2
%data,
and two anonymous referees
for their helpful comments.
This work was supported in part by grants to D.~Lin
and P.~Bodenheimer and
NASA Grant AST-9618803 
and NSF Grant NAG5-6310 to C.F.

%\clearpage

\figcaption{
Images of the
TW~Hya disk and 
PSF$-$PSF residuals.
Panel~(a) shows our
final (1620~second) image.
This 
is
a stack of 6~coadds; each coadd is a 120~or 300~second
TW~Hya frame with a 20~second PSF~star frame
scaled, shifted, and subtracted from it.
We have applied a linear stretch from %0~mJy/arcsec$^2$
0~(black) to
0.67~(white)~mJy/arcsec$^2$ %(white)
to emphasize the outer radii.
The central black circle represents our coronagraphic
mask and was added in processing; its diameter
is equal to the half-power
point of 
the Gaussian focal plane mask.
The black circle in the lower left is
a faint point source in the PSF
frames;
this point source
is at least 10~magnitudes fainter than
the 
PSF star %(SAO~202496)
and is therefore completely negligible in 
the PSF~subtractions.
Panel~(b) 
shows two H~band, 20~second PSF~images 
subtracted from each other,
and illustrates typical residuals in
the scaling and subtraction.
The linear stretch in this panel is identical
to that in panel~(a).
Panel~(c) shows 
the
I~band WFPC2~image~(K00)
with CoCo H~band contours (from panel~(a)) overlain.
The WFPC2~data are shown in logarithmic
stretch~(K00), and the CoCo
contours %are logarithmic, %with three
%evenly spaced contours per decade, 
correspond to 0.30,~0.65,~1.4, and~3.0~mJy/arcsec$^2$.
The grey and black striped pattern near the edges
of the rotated K00 image are artifacts of the rotation
interpolation scheme we used; the
original K00 image has North
18.5~degrees counter-clockwise from straight up
(J. Krist, pers. comm.).
All three panels have 
North up and East to the
left.
The field of view %for all three panels
is approximately 7$\farcs$04~on a side for
each panel.
%The effective pixel size 
%is 0.0275~arcsec after 
%an enlargement and rebinning
%of the central portion of the
%CoCo/NSFCAM field of view (14.08~arcsec on
%a side, 0.055~arcsec/pixel plate scale)
%and a rebinning of the WFPC2 
%data from K00.
\label{images}}

\figcaption{
Azimuthally averaged radial profiles for 
images of TW~Hya and the PSF~star, both
before PSF~subtraction, and of
the final (PSF-subtracted) TW~Hya circumstellar disk.
The upper and lower thin solid lines show radial
profiles of images of TW~Hya and the PSF~star,
respectively, before PSF~subtraction.
The PSF~star profile is scaled to be
equal to the TW~Hya profile at~0$\farcs$94.
The TW~Hya profile is distinctly different
from the PSF~profile outside of 
around~2\arcsec\ (110~AU), showing that we 
detect the circumstellar disk in the 
image of 
TW~Hya before PSF-subtraction.
We also show (thick solid line)
the 
radial profile of the 
final (PSF-subtracted) image of the circumstellar disk
(the image shown in Figure~1a).
The power law fit to the disk, which follows
an $r^{-3.3}$~power law,
is shown as the thick dashed line.
For comparison, the radial profile 
of the PSF$-$PSF residual (Figure~1b)
is shown in the thick dotted line;
this residual is quite small compared to the
TW~Hya disk detection.
\label{rad}}

\figcaption{
Spectral energy distribution for TW Hya.
The dashed line is a black-body 
for the star TW~Hya (4080~K).
The solid line
is the sum of stellar and disk fluxes,
including the best fit disk model found
with the five shown parameters varied.
All parameters except~$p$
%$M_{disk}$,~$R_0$,~$q$, and~$\beta$
are well constrained by the multi-wavelength
fluxes and near infrared data.
%$p$ is less well constrained.
Asterisks are near-infrared (JHK) fluxes
from \citet{webb}; 
closed circles are mid-infrared
fluxes from \citet{rayjay99b};
open circles
are IRAS fluxes \citep{iras};
squares are 0.8~and 1.1~mm data
from 
\citet{weintraub};
and the triangle is the 7~mm
flux density measurement from \citet{wilner}.
Error bars are small vertical lines
shown for each data point.
\label{sed}}

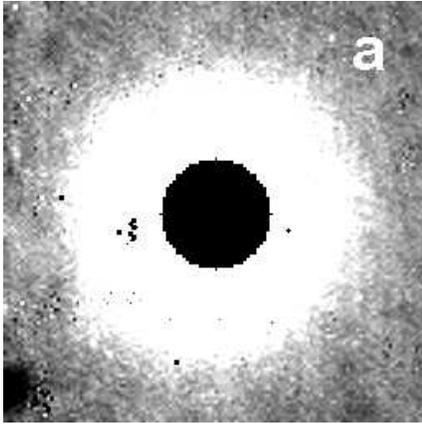
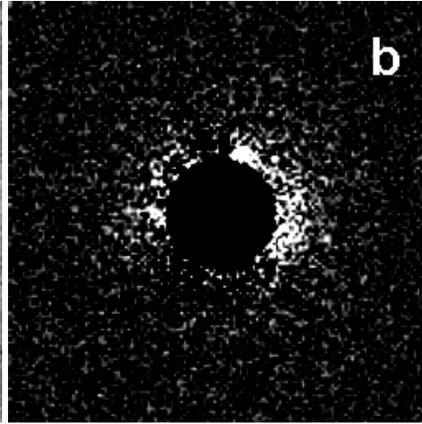
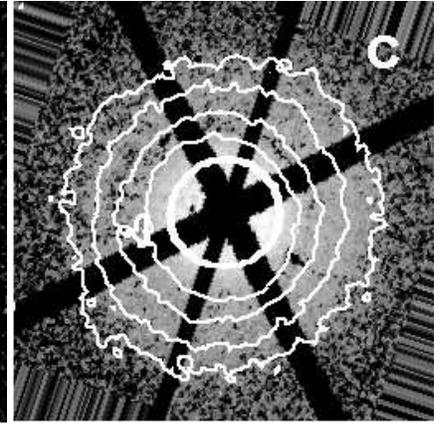
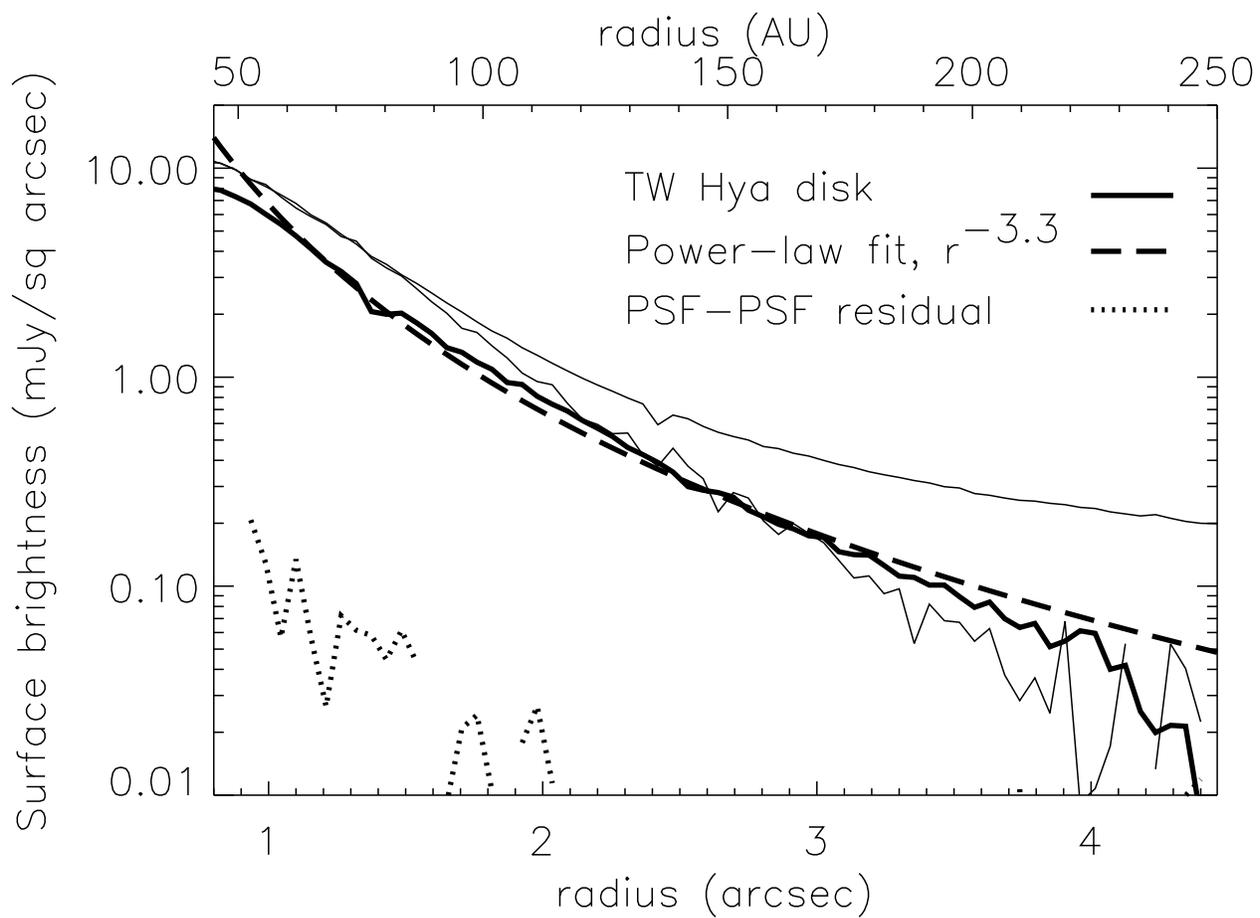
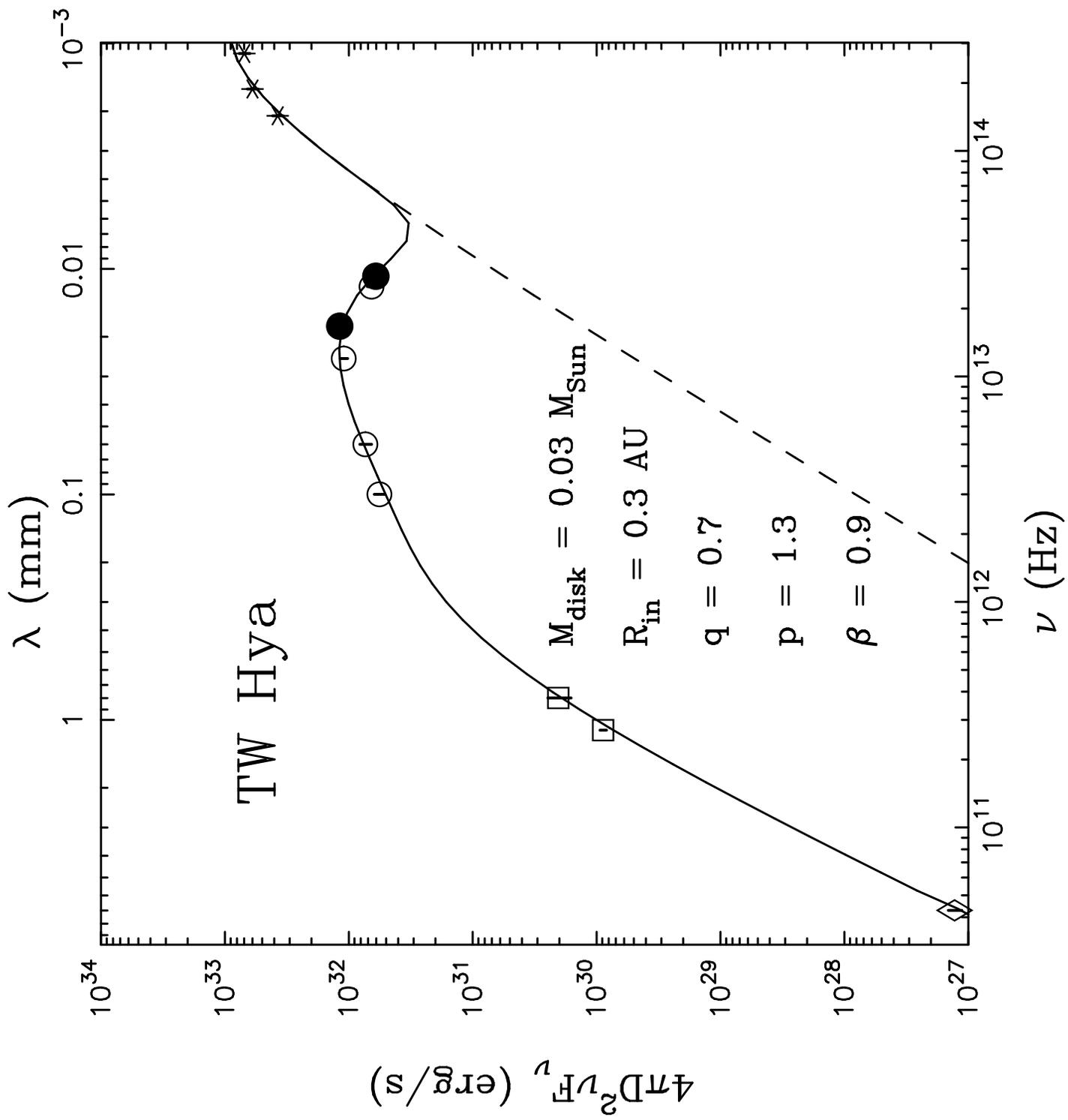
\end{document}